\documentclass[12pt]{iopart}

\usepackage{iopams}  

\newcommand{\beq}{\begin{equation}}
\newcommand{\eeq}{\end{equation}}
\newcommand{\beqs}{\begin{eqnarray}}
\newcommand{\eeqs}{\end{eqnarray}}

\begin{document}

\title{Spanning Trees on Lattices and Integration Identities} 

\author{Shu-Chiuan Chang and Wenya Wang}

\address{Department of Physics \\
National Cheng Kung University \\
Tainan 70101, Taiwan}

\begin{abstract}

For a lattice $\Lambda$ with $n$ vertices and dimension $d$ equal or higher than two, the number of spanning trees $N_{ST}(\Lambda)$ grows asymptotically as $\exp(n z_\Lambda)$ in the thermodynamic limit. We present exact integral expressions for the asymptotic growth constant $z_\Lambda$ for spanning trees on several lattices. By taking different unit cells in the calculation, many integration identities can be obtained. We also give $z_{\Lambda (p)}$ on the homeomorphic expansion of $k$-regular lattices with $p$ vertices inserted on each edge.

\end{abstract}

\maketitle

\section{Introduction}

A problem of fundamental interest in mathematics
and physics is the enumeration of the number of spanning trees on the graph
$G$, $N_{ST}(G)$.  This number can be calculated in several ways, including as
a determinant of the Laplacian matrix of $G$ and as a special case of the Tutte
polynomial of $G$ \cite{bbook}.  A previous study on the enumeration of spanning trees and the calculation of their asymptotic growth constants was carried out in Ref. \cite{sw}. In that work, closed-form integrals for these quantities were given.  
In this paper we shall present exact integrals for the asymptotic growth constant for spanning trees on five Archimedean lattices and the diamond structure. The hypercubic lattices with both nearest-neighbor and next-nearest-neighbor edges will also be considered. We shall show how integration identities can be obtained with different choices of unit cells in the calculation. We shall also derive the asymptotic growth constant for the homeomorphic expansion of a regular lattice.

\section{Background and Method}
\label{sectionmethod}

We briefly recall some definitions and background on spanning trees and the
calculation method that we use \cite{bbook,fh}.
Let $G=(V,E)$ denote a connected graph (without loops) with vertex (site) and
edge (bond) sets $V$ and $E$.  Let $n=v(G)=|V|$ be the number of vertices and
$e(G)=|E|$ the number of edges in $G$.  A spanning subgraph $G^\prime$ is a
subgraph of $G$ with $v(G^\prime) = |V|$, and a tree is a connected subgraph
with no circuits.  A spanning tree is a spanning subgraph of $G$ that is a tree
and hence $e(G') = n-1$. The degree or coordination number $k_i$ of a vertex
$v_i \in V$ is the number of edges attached to it.  Two
vertices are adjacent if they are connected by an edge.
The adjacency matrix $A(G)$ of $G$ is the $n \times n$ matrix with elements $A(G)_{ij}=1$ if $v_i$ and
$v_j$ are adjacent and zero otherwise. The Laplacian matrix $Q(G)$ is the $n
\times n$ matrix with element $Q(G)_{ij}=k_i\delta_{ij}-A(G)_{ij}$.  One of the
eigenvalues of $Q(G)$ is always zero; let us denote the rest as $\lambda(G)_i$,
$1 \le i \le n-1$.  A basic theorem is that $N_{ST}(G) =
(1/n)\prod_{i=1}^{n-1} \lambda(G)_i$ \cite{bbook}.    
For $d$-dimensional lattices $\Lambda$ with $d \ge 2$ in
the thermodynamic limit, $N_{ST}(\Lambda)$ grows exponentially with $n$ as 
$n \to \infty$; that is, there exists a constant $z_\Lambda$ such that 
$N_{ST}(\Lambda) \sim \exp(n z_\Lambda)$ as $n \to \infty$.  The constant 
describing this exponential growth is thus given by 
\beq
z_{\Lambda} = \lim_{n \to \infty} n^{-1} \ln N_{ST}(\Lambda)  
\label{zdef}
\eeq
where $\Lambda$, when used as a subscript in this manner, implicitly refers to
the thermodynamic limit of the lattice $\Lambda$. If each pairs of neighboring vertices are connected by $p$-fold edges instead of just one edge, the number of spanning trees should be multiplied by the factor $p^{n-1}$ and the resultant asymptotic growth constant increases by the factor $\ln p$. Henceforth, we only consider simple graphs without multiple edges. A regular $d$-dimensional
lattice is comprised of repeated unit cells, each containing $\nu$ vertices.
Define $a(\tilde n,\tilde n')$ as the $\nu \times \nu$ matrix describing the
adjacency of the vertices of the unit cells $\tilde
n$ and $\tilde n'$, the elements of which are given by $a(\tilde n,\tilde
n')_{ij}=1$ if $v_i \in \tilde n$ is adjacent to $v_j \in \tilde n'$ and zero
otherwise. Assuming that a given lattice has periodic boundary conditions, and using the resultant translational symmetry, we have $a(\tilde n, \tilde n')= a(\tilde n- \tilde n')$, and we can therefore write $a(\tilde n)=a(\tilde n_1,\cdots,\tilde n_d)$.  In Ref. \cite{sw} a method was derived to calculate
$N_{ST}(\Lambda)$ and $z_\Lambda$ in terms of a matrix $M$ which is
determined by these $a(\tilde n)$.  For a $d$-dimensional lattice which is $k$-regular that each of its vertices has the same degree $k$,
define
\beq
M(\theta_1,\cdots,\theta_d) = k \cdot 1 - \sum_{\tilde n} 
a(\tilde n) e^{i \tilde n \cdot \Theta } 
\label{mmatrix}
\eeq
where in this equation $1$ is the unit matrix and $\Theta$ stands for the
$d$-dimensional vector $(\theta_1,\cdots,\theta_d)$.  Then \cite{sw} 
\beq
z_\Lambda =  {1\over \nu }\int_{-\pi}^\pi 
\biggl [ \prod_{j=1}^d {d\theta_j \over {2\pi}} \biggr ] 
\ln[D(\theta_1,\cdots,\theta_d)] 
\label{zint}
\eeq
where $D(\theta_1,\cdots,\theta_d)=\det (M(\theta_1,\cdots,\theta_d))$ is the determinant of the matrix.

An Archimedean lattice is a uniform tiling of the plane by regular polygons in
which all vertices are equivalent \cite{grunbaum,shrocktsai97}.  Such a lattice can be defined by the
ordered sequence of polygons that one traverses in making a complete circuit
around the local neighborhood of any vertex.  This is indicated by the notation
$\Lambda = (\prod_i p_i^{a_i})$, meaning that in this circuit, the regular
$p_i$-sided polygon occurs contiguously $a_i$ times. All eleven Archimedean tilings are $k$-regular. 

The number of spanning trees is the same for a planar graph $G$ and its dual $G^*$, and the number of the vertices of $G^*$ is given by the Euler relation $n^*=|E|-n+1$. Because $|E|/n=k/2$ for a $k$-regular graph $G_k$, the asymptotic growth constants of $G_k$ and $G_k^*$ satisfy \cite{sw}
\beq
z_{G_k^*} = \frac{z_{G_k}}{k/2-1} \ .
\label{zdual}
\eeq 
Consequently, $z_{G_3^*} = 2z_{G_3}$, $z_{G_4^*} = z_{G_4}$, $z_{G_5^*} = 2z_{G_5}/3$ and $z_{G_6^*} = z_{G_6}/2$.

For a $k$-regular graph $G_k$, a general upper bound is $z_{G_k} \le \ln k$.  A stronger upper bound for a $k$-regular graph $G_k$ with
$k \ge 3$ can be obtained from the bound \cite{mckay,chungyau}
\beq
N_{ST}(G_k) \le \Biggl ( \frac{2\ln n}{n k \ln k} \Bigg) (b_k)^n
\label{nmckay}
\eeq
where
\beq
b_k = \frac{(k-1)^{k-1}}{[k(k-2)]^{\frac{k}{2}-1}} \ . 
\label{ck}
\eeq
With eq. (\ref{zdef}), this then yields \cite{sw} 
\beq
z_{G_k} \leq \ln(b_k) \ . 
\label{mcybound}
\eeq
It is of interest to see how close the exact results are to these upper
bounds.  For this purpose, we define the ratio
\beq
r_{G_k} = \frac{z_{G_k}}{\ln b_k} \ .
\label{rupper}
\eeq

Two graph $G$ and $G^\prime$ are homeomorphic to each other if one of them, say $G^\prime$, can be obtained from the other, $G$, by successive insertions of degree-2 vertices on edges of $G$ \cite{fh}. This process is called homeomorphic expansion. We shall denote a lattice $\Lambda$ with $p$ vertices inserted on each edge as $\Lambda(p)$.

\section{Asymptotic Growth Constants}

The asymptotic growth constants $z_\Lambda$ for many $d \ge 2$ lattices have been considered by several authors \cite{sw,temperley,wu77,tzengwu,std}. For certain two-dimensional lattices, $z_\Lambda$ can be expressed in terms of known functions. For instances, $z_{sq}=4C/\pi$, $z_{tri}=(\ln3)/2+(6/\pi)Ti_2(1/\sqrt{3})$ \cite{glasserwu,chenwu}, and $z_{(4.8.8)}=C/\pi + (\ln(\sqrt{2}-1))/2 + (Ti_2(3+2\sqrt{2}))/\pi$ \cite{std}, where $C$ is the Catalan constant and $Ti_2$ is the inverse tangent integral function \cite{invtan}. While $z_{hc}=z_{tri}/2$ due to the duality between honeycomb and triangular lattices, it is non-trivial to have the relations $z_{kag}=(z_{tri}+\ln6)/3$ and $z_{(3.12.12)}=(z_{tri}+\ln(15))/6$ given in \cite{sw}. In this section, we study the following five Archimedean lattices, $(3^3.4^2)$, $(3^2.4.3.4)$, $(3^4.6)$, $(3.4.6.4)$ and $(4.6.12)$, that have not been considered before. Most of the calculations done so far have been performed for lattices with nearest neighbor edges. The consideration can be extended to lattices with edges connecting farther vertices. We shall take the $d$-dimensional hypercubic lattice with both nearest-neighbor and next-nearest-neighbor edges for illustration.

\subsection{$(3^3.4^2)$ Lattice}

\begin{figure}[htbp]
\unitlength 1mm \hspace*{5mm}
\begin{picture}(140,50)
\multiput(0,10)(0,10){4}{\line(1,0){60}}
\multiput(10,0)(10,0){5}{\line(0,1){50}}
\multiput(0,20)(10,0){6}{\line(1,-1){10}}
\multiput(0,40)(10,0){6}{\line(1,-1){10}}
\multiput(30,10)(0,10){2}{\circle*{2}}
\put(28,22){\makebox(0,0){\small 1}}
\put(28,8){\makebox(0,0){\small 2}}
\put(30,-5){\makebox(0,0){$(a)$}}

\multiput(80,10)(0,10){4}{\line(1,0){60}}
\multiput(90,0)(10,0){5}{\line(0,1){50}}
\multiput(80,10)(20,0){3}{\line(1,-1){10}}
\multiput(80,30)(20,0){3}{\line(1,-1){10}}
\multiput(80,50)(20,0){3}{\line(1,-1){10}}
\multiput(90,10)(20,0){3}{\line(1,1){10}}
\multiput(90,30)(20,0){3}{\line(1,1){10}}
\multiput(90,20)(0,10){2}{\circle*{2}}
\multiput(100,20)(0,10){2}{\circle*{2}}
\put(98,28){\makebox(0,0){\small 1}}
\put(92,28){\makebox(0,0){\small 2}}
\put(92,22){\makebox(0,0){\small 3}}
\put(98,22){\makebox(0,0){\small 4}}
\put(110,-5){\makebox(0,0){$(b)$}}
\end{picture}

\vspace*{10mm}

\begin{picture}(140,50)
\multiput(0,10)(0,10){4}{\line(1,0){70}}
\multiput(10,0)(30,0){2}{\line(0,1){40}}
\multiput(30,10)(30,0){2}{\line(0,1){40}}
\multiput(20,0)(0,30){2}{\line(0,1){20}}
\multiput(50,0)(0,30){2}{\line(0,1){20}}
\multiput(0,10)(0,10){3}{\line(1,1){10}}
\multiput(30,10)(0,10){3}{\line(1,1){10}}
\multiput(60,10)(0,10){3}{\line(1,1){10}}
\multiput(10,0)(0,10){2}{\line(1,1){10}}
\multiput(40,0)(0,10){2}{\line(1,1){10}}
\multiput(20,30)(0,10){2}{\line(1,1){10}}
\multiput(50,30)(0,10){2}{\line(1,1){10}}
\multiput(20,20)(30,0){2}{\line(1,-1){10}}
\multiput(10,40)(30,0){2}{\line(1,-1){10}}
\multiput(10,20)(10,0){3}{\circle*{2}}
\multiput(10,30)(10,0){3}{\circle*{2}}
\put(28,28){\makebox(0,0){\small 1}}
\put(18,28){\makebox(0,0){\small 2}}
\put(12,28){\makebox(0,0){\small 3}}
\put(12,22){\makebox(0,0){\small 4}}
\put(18,22){\makebox(0,0){\small 5}}
\put(28,22){\makebox(0,0){\small 6}}
\put(35,-5){\makebox(0,0){$(c)$}}

\multiput(90,10)(0,10){4}{\line(1,0){60}}
\multiput(100,20)(10,0){2}{\line(0,1){10}}
\multiput(130,20)(10,0){2}{\line(0,1){10}}
\multiput(115,0)(10,0){2}{\line(0,1){10}}
\multiput(95,0)(50,0){2}{\line(0,1){10}}
\multiput(115,40)(10,0){2}{\line(0,1){10}}
\multiput(95,40)(50,0){2}{\line(0,1){10}}
\multiput(90,20)(10,0){2}{\line(1,-2){5}}
\multiput(120,20)(10,0){2}{\line(1,-2){5}}
\multiput(105,40)(10,0){2}{\line(1,-2){5}}
\multiput(135,40)(10,0){2}{\line(1,-2){5}}
\multiput(105,10)(10,0){2}{\line(1,2){5}}
\multiput(135,10)(10,0){2}{\line(1,2){5}}
\multiput(90,30)(10,0){2}{\line(1,2){5}}
\multiput(120,30)(10,0){2}{\line(1,2){5}}
\multiput(110,20)(10,0){3}{\circle*{2}}
\multiput(110,30)(10,0){3}{\circle*{2}}
\put(128,28){\makebox(0,0){\small 1}}
\put(118,28){\makebox(0,0){\small 2}}
\put(112,28){\makebox(0,0){\small 3}}
\put(112,22){\makebox(0,0){\small 4}}
\put(118,22){\makebox(0,0){\small 5}}
\put(128,22){\makebox(0,0){\small 6}}
\put(120,-5){\makebox(0,0){$(d)$}}
\end{picture}

\vspace*{10mm}

\begin{picture}(100,80)
\multiput(0,20)(0,30){2}{\line(1,0){10}}
\multiput(20,20)(0,30){2}{\line(1,0){30}}
\multiput(60,20)(0,30){2}{\line(1,0){30}}
\multiput(0,0)(40,0){3}{\line(0,1){20}}
\multiput(30,0)(40,0){2}{\line(0,1){20}}
\multiput(0,50)(40,0){3}{\line(0,1){30}}
\multiput(30,50)(40,0){2}{\line(0,1){30}}
\multiput(10,10)(10,0){2}{\line(0,1){50}}
\multiput(50,10)(10,0){2}{\line(0,1){50}}
\put(90,10){\line(0,1){50}}
\multiput(0,10)(40,0){2}{\line(1,0){30}}
\multiput(0,60)(40,0){2}{\line(1,0){30}}
\multiput(80,10)(0,50){2}{\line(1,0){20}}
\multiput(10,30)(40,0){3}{\line(1,0){10}}
\multiput(10,40)(40,0){3}{\line(1,0){10}}
\multiput(30,70)(40,0){2}{\line(1,0){10}}
\multiput(50,10)(0,10){6}{\circle*{2}}
\multiput(60,10)(0,10){6}{\circle*{2}}
\put(62,52){\makebox(0,0){\small 1}}
\put(62,62){\makebox(0,0){\small 2}}
\put(48,62){\makebox(0,0){\small 3}}
\put(48,52){\makebox(0,0){\small 4}}
\put(48,42){\makebox(0,0){\small 5}}
\put(62,42){\makebox(0,0){\small 6}}
\put(62,32){\makebox(0,0){\small 7}}
\put(48,32){\makebox(0,0){\small 8}}
\put(48,22){\makebox(0,0){\small 9}}
\put(48,12){\makebox(0,0){\small 10}}
\put(62,12){\makebox(0,0){\small 11}}
\put(62,22){\makebox(0,0){\small 12}}
\put(40,-5){\makebox(0,0){$(e)$}}
\end{picture}

\vspace*{5mm}
\caption{\footnotesize{(a) The $(3^3.4^2)$ lattice. (b) The $(3^2.4.3.4)$ lattice. (c) The $(3^4.6)$ lattice. (d) The $(3.4.6.4)$ lattice. (e) The $(4.6.12)$ lattice. Vertices within a unit cell are labeled.}} 
\label{archimedean}
\end{figure}
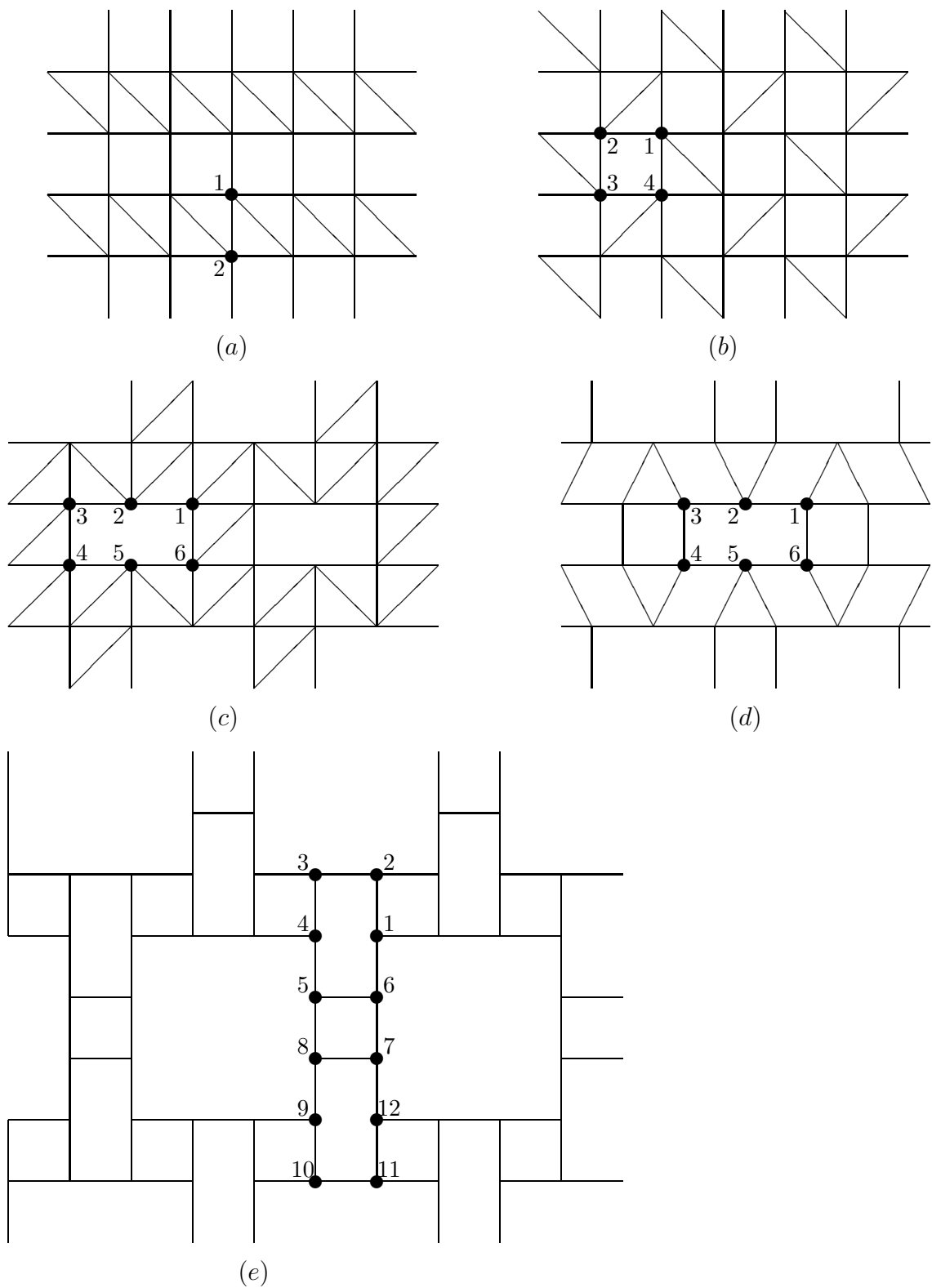

The $(3^3.4^2)$ lattice can be constructed by starting with the square lattice and adding a diagonal edge connecting the vertices in, say, the upper left to the lower right corners of each square in every other row as shown in Fig. \ref{archimedean} (a). The simplest unit cell for this lattice contains two vertices $\nu_{(3^3.4^2)}=2$ as labeled in Fig. \ref{archimedean} (a).
We have
\beq
M_{(3^3.4^2)}(\theta_1,\theta_2) = \left( \begin{array}{cc}
5 - 2\cos{\theta_1} & -1-e^{i\theta_1}-e^{i\theta_2} \\
-1-e^{-i\theta_1}-e^{-i\theta_2} & 5-2\cos{\theta_1} 
\end{array} \right ) \ .
\eeq
The determinant can be calculated to be
\beq
D_{(3^3.4^2)}(\theta_1,\theta_2) = 22 - 22\cos \theta_1 + 4\cos^2 \theta_1 - 2\cos \theta_2 - 2\cos(\theta_1-\theta_2)
\eeq
such that
\beqs
z_{(3^3.4^2)} & = & \frac{1}{2} \int_{-\pi}^{\pi} \frac{d\theta_1}{2\pi} \int_{-\pi}^{\pi} \frac{d\theta_2}{2\pi}
\ln \Big [ D_{(3^3.4^2)}(\theta_1,\theta_2) \Big ] \cr\cr
& = & \frac{1}{2} \int_{-\pi}^{\pi} \frac{d\theta}{2\pi} \ln \Big [ 11-11\cos \theta +2\cos^2 \theta \cr & & \qquad + \sqrt{(1-\cos \theta)(7-2\cos \theta)(17-13\cos \theta +2\cos^2 \theta)} \Big ] \cr\cr
& = & 1.406925832
\eeqs
where one integration is carried out followed by a numerical evaluation of the remaining integral.

\subsection{$(3^2.4.3.4)$ Lattice}

The $(3^2.4.3.4)$ lattice is shown in Fig. \ref{archimedean} (b) which has the structure of a net of squares. Taking each square as a unit cell with $\nu_{(3^2.4.3.4)}=4$, we have
\beqs
& & M_{(3^2.4.3.4)}(\theta_1,\theta_2) \cr\cr
& = & \left( \begin{array}{cccc}
5 & -1-e^{i\theta_1} & -e^{i\theta_1} & -1-e^{i\theta_2} \\
-1-e^{-i\theta_1} & 5 & -1-e^{i\theta_2} & -e^{i\theta_2} \\
-e^{-i\theta_1} & -1-e^{-i\theta_2} & 5 & -1-e^{-i\theta_1} \\
-1-e^{-i\theta_2} & -e^{-i\theta_2} & -1-e^{i\theta_1} & 5
\end{array} \right ) \ .
\eeqs
The determinant can be calculated to be
\beqs
D_{(3^2.4.3.4)}(\theta_1,\theta_2) & = & 336 - 144(\cos \theta_1 + \cos \theta_2) + 4(\cos^2 \theta_1 + \cos^2 \theta_2) \cr\cr
& & - 56\cos \theta_1 \cos \theta_2
\eeqs
such that
\beqs
& & z_{(3^2.4.3.4)} = \frac{1}{4} \int_{-\pi}^{\pi} \frac{d\theta_1}{2\pi} \int_{-\pi}^{\pi} \frac{d\theta_2}{2\pi}
\ln \Big [ D_{(3^2.4.3.4)}(\theta_1,\theta_2) \Big ] \cr\cr
& = & \ln 2 + \frac{1}{4} \int_{-\pi}^{\pi} \frac{d\theta_1}{2\pi} \int_{-\pi}^{\pi} \frac{d\theta_2}{2\pi}
\ln \Big [ 6 + 4\cos(\frac{\theta_1}{2})\cos(\frac{\theta_2}{2}) - \cos^2(\frac{\theta_1}{2}) - \cos^2(\frac{\theta_2}{2}) \Big ] \cr & & + \frac{1}{4} \int_{-\pi}^{\pi} \frac{d\theta_1}{2\pi} \int_{-\pi}^{\pi} \frac{d\theta_2}{2\pi} \ln \Big [ 6 - 4\cos(\frac{\theta_1}{2})\cos(\frac{\theta_2}{2}) - \cos^2(\frac{\theta_1}{2}) - \cos^2(\frac{\theta_2}{2}) \Big ] \cr\cr
& = & 1.410855646 \ .
\eeqs

\subsection{$(3^4.6)$ Lattice}

The $(3^4.6)$ lattice is shown in Fig. \ref{archimedean} (c) which has the structure of a net of hexagons. Taking each hexagon as a unit cell with $\nu_{(3^4.6)}=6$, we have
\beqs
& & M_{(3^4.6)}(\theta_1,\theta_2) \cr\cr
& = & \left( \begin{array}{cccccc}
5 & -1 & -e^{i\theta_1} & -e^{i(\theta_1+\theta_2)} & -e^{i(\theta_1+\theta_2)} & -1 \\
-1 & 5 & -1 & -e^{i(\theta_1+\theta_2)} & -e^{i\theta_2} & -e^{i\theta_2} \\
-e^{-i\theta_1} & -1 & 5 & -1 & -e^{i\theta_2} & -e^{-i\theta_1} \\
-e^{-i(\theta_1+\theta_2)} & -e^{-i(\theta_1+\theta_2)} & -1 & 5 & -1 & -e^{-i\theta_1} \\
-e^{-i(\theta_1+\theta_2)} & -e^{-i\theta_2} & -e^{-i\theta_2} & -1 & 5 & -1 \\
-1 & -e^{-i\theta_2} & -e^{i\theta_1} & -e^{i\theta_1} & -1 & 5
\end{array} \right ) \cr & &
\eeqs
and the determinant $D_{(3^4.6)}(\theta_1,\theta_2)$ is given by $f(5032,1624,56)$, where we define the function
\beqs
f(a,b,c) & = & a - b[\cos \theta_1 +\cos \theta_2 + \cos(\theta_1-\theta_2)] \cr & & - c[\cos \theta_1 \cos \theta_2 + (\cos \theta_1 +\cos \theta_2) \cos(\theta_1-\theta_2)] \cr & & + 8\cos \theta_1 \cos \theta_2 \cos(\theta_1-\theta_2) \ .
\eeqs
Thus,
\beq
z_{(3^4.6)} = \frac{1}{6} \int_{-\pi}^{\pi} \frac{d\theta_1}{2\pi} \int_{-\pi}^{\pi} \frac{d\theta_2}{2\pi} \ln \Big [ f(5032,1624,56) \Big ] = 1.392023563 \ .
\eeq
This result applies to both enantiomorphic forms of the $(3^4.6)$ lattice.

\subsection{$(3.4.6.4)$ Lattice}

The $(3.4.6.4)$ lattice is shown in Fig. \ref{archimedean} (d) which also has the structure of a net of hexagons. Taking each hexagon as a unit cell with $\nu_{(3.4.6.4)}=6$, we have
\beqs
& & M_{(3.4.6.4)}(\theta_1,\theta_2) \cr\cr & = & \left( \begin{array}{cccccc}
4 & -1 & -e^{i\theta_1} & 0 & -e^{i(\theta_1+\theta_2)} & -1 \\
-1 & 4 & -1 & -e^{i(\theta_1+\theta_2)} & 0 & -e^{i\theta_2} \\
-e^{-i\theta_1} & -1 & 4 & -1 & -e^{i\theta_2} & 0 \\
0 & -e^{-i(\theta_1+\theta_2)} & -1 & 4 & -1 & -e^{-i\theta_1} \\
-e^{-i(\theta_1+\theta_2)} & 0 & -e^{-i\theta_2} & -1 & 4 & -1 \\
-1 & -e^{-i\theta_2} & 0 & -e^{i\theta_1} & -1 & 4
\end{array} \right ) \ . \cr & &
\eeqs
The determinant $D_{(3.4.6.4)}(\theta_1,\theta_2)$ is given by $f(1144,376,8)$ such that
\beq
z_{(3.4.6.4)} = \frac{1}{6} \int_{-\pi}^{\pi} \frac{d\theta_1}{2\pi} \int_{-\pi}^{\pi} \frac{d\theta_2}{2\pi}
\ln \Big [ f(1144,376,8) \Big ] = 1.144801124 \ .
\eeq

\subsection{$(4.6.12)$ Lattice}

The $(4.6.12)$ lattice is shown in Fig. \ref{archimedean} (e), where every two neighboring hexagons is taken as a unit cell with $\nu_{(4.6.12)}=12$. We have
\beqs
& & M_{(4.6.12)}(\theta_1,\theta_2) \cr & & = \left( \begin{array}{cccccccccccc}
3 & -1 & 0 & 0 & 0 & -1 & 0 & 0 & 0 & -e^{i\theta_1} & 0 & 0 \\
-1 & 3 & -1 & 0 & 0 & 0 & 0 & 0 & -e^{i\theta_1} & 0 & 0 & 0 \\
0 & -1 & 3 & -1 & 0 & 0 & 0 & 0 & 0 & 0 & 0 & -e^{i\theta_2} \\
0 & 0 & -1 & 3 & -1 & 0 & 0 & 0 & 0 & 0 & -e^{i\theta_2} & 0 \\
0 & 0 & 0 & -1 & 3 & -1 & 0 & -1 & 0 & 0 & 0 & 0 \\
-1 & 0 & 0 & 0 & -1 & 3 & -1 & 0 & 0 & 0 & 0 & 0 \\
0 & 0 & 0 & 0 & 0 & -1 & 3 & -1 & 0 & 0 & 0 & -1 \\
0 & 0 & 0 & 0 & -1 & 0 & -1 & 3 & -1 & 0 & 0 & 0 \\ 
0 & -e^{-i\theta_1} & 0 & 0 & 0 & 0 & 0 & -1 & 3 & -1 & 0 & 0 \\
-e^{-i\theta_1} & 0 & 0 & 0 & 0 & 0 & 0 & 0 & -1 & 3 & -1 & 0  \\
0 & 0 & 0 & -e^{-i\theta_2} & 0 & 0 & 0 & 0 & 0 & -1 & 3 & -1 \\
0 & 0 & -e^{-i\theta_2} & 0 & 0 & 0 & -1 & 0 & 0 & 0 & -1 & 3
\end{array} \right ) \ . \cr & &
\eeqs
The determinant $D_{(4.6.12)}(\theta_1,\theta_2)$ is given by $f(13432,4344,136)$ such that
\beqs
z_{(4.6.12)} & = & \frac{1}{12} \int_{-\pi}^{\pi} \frac{d\theta_1}{2\pi} \int_{-\pi}^{\pi} \frac{d\theta_2}{2\pi}
\ln \Big [ f(13432,4344,136) \Big ] = 0.7777955061 \ . \cr & &
\eeqs

\subsection{$d$-Dimensional Hypercubic Lattice with Next-Nearest-Neighbor Edges}

The $d$-dimensional hypercubic lattice, denoted as $c(d)$, with nearest-neighbor edges has been studied before in Refs. \cite{sw,tzengwu}. Now consider the hypercubic lattice with next-nearest-neighbor edges included, denoted as $c(d),nnn$. The one-dimensional circuit graph with next-nearest-neighbor edges corresponds to a strip of the triangular lattice with width two. We have
\beq
M_{c(1),nnn}(\theta) = \left( \begin{array}{cc}
4-2\cos \theta & -1-e^{i\theta} \\
-1-e^{-i\theta} & 4-2\cos \theta
\end{array} \right ) \ .
\eeq
The determinant $D_{c(1),nnn}(\theta)$ is equal to $2(1-\cos \theta)(7-2\cos \theta)$ such that 
\beq
z_{c(1),nnn} = (1/2) \ln [ (7+3\sqrt{5})/2 ] = 0.9624236501 
\eeq
which agrees with Refs. \cite{sw,ta}. For $d \ge 2$, the coordination number is given by $k_{c(d),nnn} = 2d+4{d \choose 2}=2d^2$. We have the general expression
\beqs
z_{c(d),nnn} & = & \int_{-\pi}^{\pi} \Big [ \prod_{i=1}^d \frac{d\theta_i}{2\pi} \Big ] \ln \Big ( 2d^2 -2\sum_{i=1}^d \cos \theta_i -2\sum_{i \ne j} \cos \theta_i \cos \theta_j \Big ) \cr
& & \qquad \qquad \qquad \qquad \qquad \qquad \qquad \mbox{for} \ d \ge 2 \ .
\eeqs
The square lattice with next-nearest-neighbor edges corresponds to $c(d=2),nnn$ that
\beqs
z_{sq,nnn} & = & \int_{-\pi}^{\pi} \frac{d\theta_1}{2\pi} \int_{-\pi}^{\pi} \frac{d\theta_2}{2\pi} \ln \Big [ 8 -2\cos \theta_1 -2\cos \theta_2 -4 \cos \theta_1 \cos \theta_2 \Big ] \cr\cr
& = & \int_0^{\pi} \frac{d\theta}{\pi} \ln \Big [ 4-\cos \theta +\sqrt{3(1-\cos \theta)(5+\cos \theta)} \Big ] \ .
\eeqs
An exact closed form expression for this integral can be derived.  We begin by
recasting the integral in the equivalent form.  
\beqs
z_{sq,nnn} & = & \ln 3 + \frac{4}{\pi} \int_0^{\pi/2} d\phi\ln \Big [ \sin \phi + \sqrt{1-(1/3)\sin^2 \phi} \Big ] \cr\cr
& = & \ln 3 + 4I(1/\sqrt{3})
\label{zsqnnn}
\eeqs
where
\beq
I(a) = \frac{1}{\pi} \int_0^{\pi/2} d\phi \ \ln \Bigl ( \sin\phi 
+ \sqrt{1-a^2 \sin^2\phi } \ \Bigr ) \ . 
\label{iint}
\eeq
In eq. (\ref{iint}), with no loss of generality, we take $a$ to be nonnegative.
We will give a general result for $I(a)$ with $0 \le a < 1$ and then specialize to our case $a=1/\sqrt{3}$.  First, we note that $I(1)=C/\pi$, where $C$ is the Catalan constant. Next, taking the derivative with respect to $a$ and doing the integral over $\phi$ in eq. (\ref{iint}), we get
\beq
I'(a) = \frac{a/2 - (2/\pi)\tanh^{-1}a}{(1+a^2)} \ . 
\label{iprime}
\eeq
To calculate $I(a)$, we then use $I(a)-I(0) = \int_{0}^{a} I'(x) dx$ and
observe that 
\beq
I(0) = \frac{1}{\pi} \int_{0}^{\pi/2} d\phi \ \ln (\sin \phi +1) =
-\frac{\ln 2}{2} + \frac{2C}{\pi} \ .
\label{izero}
\eeq
We also make use of the integrals 
\beq
\int_{0}^{a} \frac{x}{1+x^2} \ dx = \frac{1}{2} \ln (1+a^2)
\label{int1a}
\eeq
and
\beq
\int_{0}^{a} \frac{\tanh^{-1}x}{1+x^2} \ dx = \tan^{-1}a \tanh^{-1}a +\frac{C}{2} +\frac{\pi}{8} \ln \biggl ( \frac{1+a}{1-a} \biggr ) - 
\frac{1}{2} {\rm Ti}_2 \biggl ( \frac{1+a}{1-a} \biggr )
\eeq
to obtain
\beq
I(a) = \frac{C}{\pi} + \frac{1}{4} \ln \Big ( \frac{(1+a^2)(1-a)}{4(1+a)} \Big ) -\frac{2}{\pi} \tan^{-1}a \tanh^{-1}a
+ \frac{1}{\pi}{\rm Ti}_2 \biggl ( \frac{1+a}{1-a} \biggr ) 
\label{iabelow1}
\eeq
where Ti$_2(x)$ is the inverse tangent integral \cite{invtan}, 
\beqs
{\rm Ti}_2(x) & = & \int_0^x \frac{\tan^{-1} y}{y} \ dy \cr\cr
        & = & x[ \ {}_3F_2([1,1/2,1/2],[3/2,3/2],-x^2) \ ] \ . 
\label{ti2}
\eeqs
Evaluating our result (\ref{iabelow1}) for $I(a)$ at
$a=1/\sqrt{3}$ and substituting into eq. (\ref{zsqnnn}), we obtain the exact,
closed-form expression
\beqs
z_{sq,nnn} & = & \frac{4C}{\pi} + \ln (2 - \sqrt{3}) -\frac43 \tanh^{-1} \frac{1}{\sqrt{3}}
+ \frac{4}{\pi} {\rm Ti}_2(2+\sqrt{3} \, ) \cr\cr
& = & 1.943739360 \ .
\label{zsqnnne}
\eeqs
We note that $z_{sq}$ and $z_{tri}$ for the square and triangular lattices, respectively, can be expressed in terms of the ${\rm Ti}_2$ function
\cite{glasserwu,chenwu}. The numerical evaluations for $3 \le d \le 5$ are given by 
\beqs
z_{c(3),nnn} & = & 2.841787536 \cr
z_{c(4),nnn} & = & 3.442108 \cr
z_{c(5),nnn} & = & 3.898251 \ . 
\eeqs
With the observation that $z_{c(d),nnn} \to \ln (2d^2)$ for $d \to \infty$ and similar behavior for other $d$-dimensional lattices \cite{sw}, we conjecture that for a $d$-dimensional lattice $\Lambda$ with degree $k_d$, $z_{\Lambda}$ approaches to $\ln (k_d)$ from below in the large $d$ limit.

\section{Integral Identities}

As the method reviewed in section \ref{sectionmethod} is general, there is a freedom to specify the unit cell for the calculation. Many integral identities can be obtained with different choices of unit cells in the calculation of the spanning trees. 

\subsection{bcc Lattice}

Consider the body-centered cubic (bcc) lattice with coordination number $k_{bcc}=8$ as the first example. Although the conventional bcc cell contains two vertices, the primitive cell contains only one vertex. In terms of the unit vectors $\hat x$, $\hat y$, $\hat z$ of the Cartesian coordinate, the primitive translation vectors of the bcc lattice are $(\hat x+\hat y-\hat z)/2$, $(\hat x-\hat y+\hat z)/2$ and $(-\hat x+\hat y+\hat z)/2$. Hence,
\beq
z_{bcc} = 3\ln 2 + \int_{-\pi}^{\pi}{{d\theta_1}\over{2\pi}}
 \int_{-\pi}^{\pi}{{d\theta_2}\over{2\pi}}
 \int_{-\pi}^{\pi}{{d\theta_3}\over{2\pi}}
 \ln  \Big [ F_{bcc}(\theta_1,\theta_2,\theta_3) \Big ]
\label{zbcc}
\eeq
where 
\beq
F_{bcc}(\theta_1,\theta_2,\theta_3) = 1 - \frac14 \Big ( \cos \theta_1 + \cos \theta_2 + \cos \theta_3 + \cos(\theta_1+\theta_2+\theta_3) \Big ) \ .
\eeq
Comparing $z_{bcc}$ with the expressions given in eq.(5.2.7) of Ref. \cite{sw} and in eq.(16) of Ref. \cite{std}, we find
\beqs
& & \int_{-\pi}^{\pi}{{d\theta_1}\over{2\pi}}
 \int_{-\pi}^{\pi}{{d\theta_2}\over{2\pi}}
 \int_{-\pi}^{\pi}{{d\theta_3}\over{2\pi}}
 \ln  \Big [ F_{bcc}(\theta_1,\theta_2,\theta_3) \Big ] \cr\cr
& = &  -\frac{1}{2}\sum_{\ell=1}^\infty \frac{1}{\ell}
\Biggl (\frac{(2\ell)!}{2^{2\ell} \, (\ell !)^2} \Biggr )^3 \cr\cr 
& = & - \frac{ {}_5F_4([1,1,3/2,3/2,3/2], [2,2,2,2],1)} {2^4} \ .
\eeqs

\subsection{fcc Lattice}

Next, consider the face-centered-cubic (fcc) lattice with coordination number $k_{fcc}=12$. The conventional fcc cell contains four vertices, but the primitive cell again contains only one vertex. The primitive translation vectors of the fcc lattice are $(\hat x+\hat y)/2$, $(\hat y+\hat z)/2$ and $(\hat z+\hat x)/2$. Hence,
\beq
z_{fcc} = \ln (12) + \int_{-\pi}^{\pi}{{d\theta_1}\over{2\pi}}
 \int_{-\pi}^{\pi}{{d\theta_2}\over{2\pi}}
 \int_{-\pi}^{\pi}{{d\theta_3}\over{2\pi}}
 \ln \Big [ F_{fcc}(\theta_1,\theta_2,\theta_3) \Big ]
\label{zfcc}
\eeq
where 
\beqs
F_{fcc}(\theta_1,\theta_2,\theta_3) & = & 1 - \frac16 \Big ( \cos \theta_1 + \cos \theta_2 + \cos \theta_3 \cr\cr & & \qquad + \cos(\theta_1-\theta_2) + \cos(\theta_2-\theta_3) + \cos(\theta_3-\theta_1) \Big ) \ .
\label{Ffcc}
\eeqs
The equality of this triple integral and the triple integral given in eqs.(18) and (19) of Ref. \cite{std} is therefore established.

\subsection{Diamond Structure}

The diamond structure can be viewed as two fcc structures displaced from each other by one-quarter of a cubic diagonal. Each vertex has four nearest-neighbor, $k_{diamond}=4$, at the corners of a tetrahedron. The conventional unit cube contains eight vertices, so that $\nu_{diamond}=8$, located at $(0,0,0)$, $(\frac12,\frac12,0)$, $(\frac12,0,\frac12)$, $(0,\frac12,\frac12)$, $(\frac14,\frac14,\frac14)$, $(\frac34,\frac34,\frac14)$, $(\frac34,\frac14,\frac34)$ and $(\frac14,\frac34,\frac34)$. We have
\beqs
& & M_{diamond} \cr\cr & & =
\left(\begin{array}{cccccccc}
4 & 0 & 0 & 0 & -1 & -e^{-i(\theta_1+\theta_2)} & -e^{-i(\theta_1+\theta_3)} & -e^{-i(\theta_2+\theta_3)}  \\
0 & 4 & 0 & 0 & -1 & -1 & -e^{-i\theta_3} & -e^{-i\theta_3}  \\
0 & 0 & 4 & 0 & -1 & -e^{-i\theta_2} & -1 & -e^{-i\theta_2}   \\
0 & 0 & 0 & 4 & -1 & -e^{-i\theta_1} & -e^{-i\theta_1} & -1   \\  
-1 & -1 & -1 & -1 & 4 & 0 & 0 & 0  \\
-e^{i(\theta_1+\theta_2)} & -1 & -e^{i\theta_2} & -e^{i\theta_1} & 0 & 4 & 0 & 0  \\
-e^{i(\theta_1+\theta_3)} & -e^{i\theta_3} & -1 & -e^{i\theta_1} & 0 & 0 & 4 & 0  \\
-e^{i(\theta_2+\theta_3)} & -e^{i\theta_3} & -e^{i\theta_2} & -1 & 0 & 0 & 0 & 4
\end{array} \right) \ . \cr & &
\label{Mdia}
\eeqs
The determinant of this matrix is exactly the same as that for the fcc lattice, i.e., $12^4F(\theta_1,\theta_2,\theta_3)$ where $F(\theta_1,\theta_2,\theta_3)$ is given in eq.(19) of Ref. \cite{std}.
Therefore, 
\beq
z_{diamond} = z_{fcc}/2 \simeq 1.206459950 \ . 
\eeq
This result can be verified by taking the primitive basis with two vertices located at $(0,0,0)$ and $(\frac14,\frac14,\frac14)$. The corresponding matrix is given by
\beq
\bar M_{diamond} =
\left(\begin{array}{cc}
4 & -1-e^{-i\theta_1}-e^{-i\theta_2}-e^{-i\theta_3}  \\
-1-e^{i\theta_1}-e^{i\theta_2}-e^{i\theta_3} & 4 
\end{array} \right) 
\label{Mdia2}
\eeq
with determinant $12F_{fcc}(\theta_1,\theta_2,\theta_3)$ where $F_{fcc}(\theta_1,\theta_2,\theta_3)$ is given above in eq. (\ref{Ffcc}).

\subsection{Square Lattice}

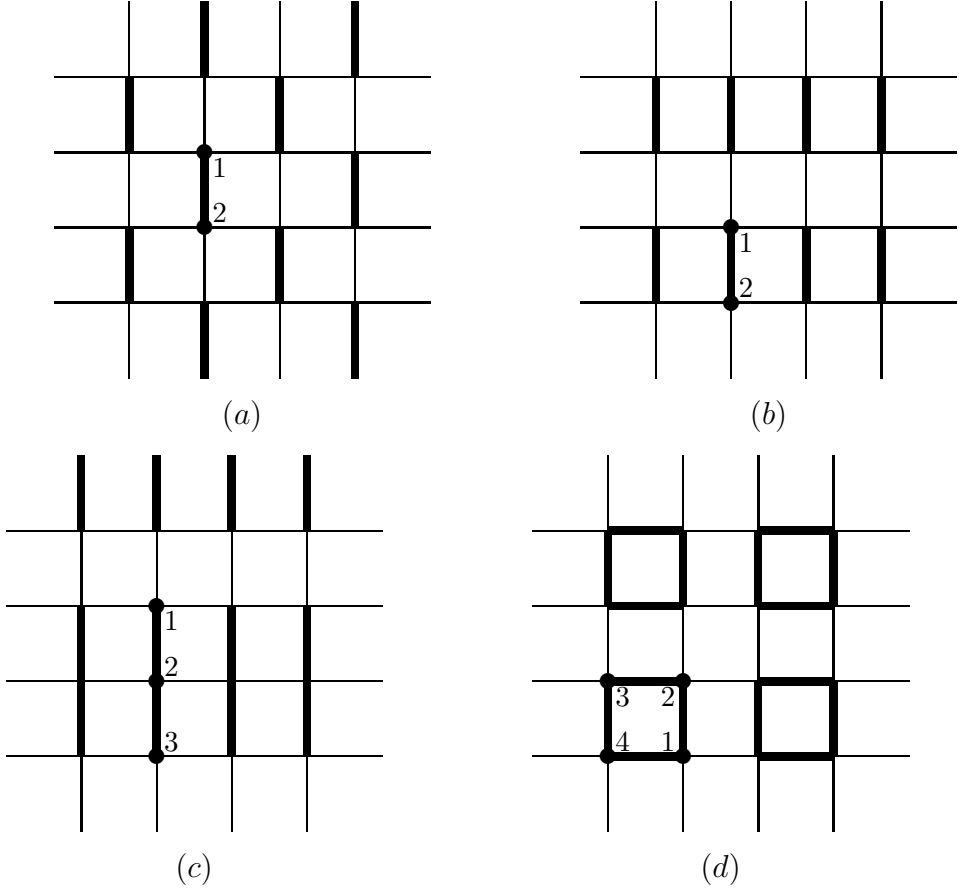
\begin{figure}[htbp]
\unitlength 1mm \hspace*{5mm}
\begin{picture}(120,50)
\multiput(0,10)(0,10){4}{\line(1,0){50}}
\multiput(10,0)(10,0){4}{\line(0,1){50}}
\multiput(20,20)(0,10){2}{\circle*{2}}
\put(22,28){\makebox(0,0){\small 1}}
\put(22,22){\makebox(0,0){\small 2}}
{\linethickness{1mm}
\multiput(20,0)(20,0){2}{\line(0,1){10}} 
\multiput(10,10)(20,0){2}{\line(0,1){10}}
\multiput(20,20)(20,0){2}{\line(0,1){10}} 
\multiput(10,30)(20,0){2}{\line(0,1){10}}
\multiput(20,40)(20,0){2}{\line(0,1){10}} }
\put(25,-5){\makebox(0,0){$(a)$}}

\multiput(70,10)(0,10){4}{\line(1,0){50}}
\multiput(80,0)(10,0){4}{\line(0,1){50}}
\multiput(90,10)(0,10){2}{\circle*{2}}
\put(92,18){\makebox(0,0){\small 1}}
\put(92,12){\makebox(0,0){\small 2}}
{\linethickness{1mm}
\multiput(80,10)(10,0){4}{\line(0,1){10}} 
\multiput(80,30)(10,0){4}{\line(0,1){10}} }
\put(95,-5){\makebox(0,0){$(b)$}}
\end{picture}

\vspace*{10mm}

\begin{picture}(120,50)
\multiput(0,10)(0,10){4}{\line(1,0){50}}
\multiput(10,0)(10,0){4}{\line(0,1){50}}
\multiput(20,10)(0,10){3}{\circle*{2}}
\put(22,28){\makebox(0,0){\small 1}}
\put(22,22){\makebox(0,0){\small 2}}
\put(22,12){\makebox(0,0){\small 3}}
{\linethickness{1mm}
\multiput(10,10)(10,0){4}{\line(0,1){20}} 
\multiput(10,40)(10,0){4}{\line(0,1){10}} }
\put(25,-5){\makebox(0,0){$(c)$}}

\multiput(70,10)(0,10){4}{\line(1,0){50}}
\multiput(80,0)(10,0){4}{\line(0,1){50}}
\multiput(80,10)(0,10){2}{\circle*{2}}
\multiput(90,10)(0,10){2}{\circle*{2}}
\put(88,12){\makebox(0,0){\small 1}}
\put(88,18){\makebox(0,0){\small 2}}
\put(82,18){\makebox(0,0){\small 3}}
\put(82,12){\makebox(0,0){\small 4}}
{\linethickness{1mm}
\multiput(80,10)(10,0){4}{\line(0,1){10}} 
\multiput(80,30)(10,0){4}{\line(0,1){10}}
\multiput(80,10)(0,10){4}{\line(1,0){10}} 
\multiput(100,10)(0,10){4}{\line(1,0){10}} }
\put(95,-5){\makebox(0,0){$(d)$}}
\end{picture}

\vspace*{5mm}
\caption{\footnotesize{Different unit cells for the square lattice. (a) The choice for the $bcc(2)$ lattice with $\nu=2$. (b) The other choice with $\nu=2$. (c) A choice with $\nu=3$. (d) A choice with $\nu=4$. Vertices within a unit cell are labeled and edges within each unit cell are connected by thick lines.}} 
\label{sqfig}
\end{figure}

Let us take the square lattice as another illustration. It has been pointed out in Ref. \cite{sw} that $d=2$ case of the generalized body-centered-cubic lattice $bcc(d)$ is the square lattice. The unit cell for $bcc(2)$ with $\nu=2$ is shown in Fig. \ref{sqfig} (a). If one takes the unit cell containing two vertices as shown in Fig.\ref{sqfig} (b), the corresponding matrix is
\beq
M_{sq,2} =
\left(\begin{array}{cc}
4-2\cos \theta_1 & -1-e^{i\theta_2}  \\
-1-e^{-i\theta_2} & 4-2\cos \theta_1 
\end{array} \right) 
\label{Msq2}
\eeq
with determinant
\beq
D_{sq,2}(\theta_1,\theta_2) = 14-16\cos \theta_1 +4\cos^2 \theta_1 -2\cos \theta_2 \ .
\eeq
A possible choice of the unit cell containing three vertices for the square lattice is shown in Fig. \ref{sqfig} (c). The corresponding matrix is
\beq
M_{sq,3} =
\left(\begin{array}{ccc}
4-2\cos \theta_1 & -1 & -e^{i\theta_2}  \\
-1 & 4-2\cos \theta_1 & -1  \\
-e^{-i\theta_2} & -1 & 4-2\cos \theta_1 
\end{array} \right) 
\label{Msq3}
\eeq
with determinant
\beq
D_{sq,3}(\theta_1,\theta_2) =
52-90\cos \theta_1 +48\cos^2 \theta_1-8\cos^3 \theta_1-2\cos \theta_2 \ .
\eeq
A possible choice of the unit cell containing four vertices for the square lattice is shown in Fig. \ref{sqfig} (d). The corresponding matrix is
\beq
M_{sq,4} =
\left(\begin{array}{cccc}
4 & -1-e^{-i\theta_2} & 0 & -1-e^{i\theta_1}  \\
-1-e^{i\theta_2} & 4 & -1-e^{i\theta_1} & 0  \\
0 & -1-e^{-i\theta_1} & 4 & -1-e^{i\theta_2} \\
-1-e^{-i\theta_1} & 0 & -1-e^{-i\theta_2} & 4 
\end{array} \right) 
\label{Msq4}
\eeq
with determinant
\beq
D_{sq,4}(\theta_1,\theta_2) =
128-64(\cos \theta_1 +\cos \theta_2)+4(\cos \theta_1 -\cos \theta_2)^2 \ .
\eeq
All of these considerations give the same $z_{sq}$ for the square lattice \cite{temperley,wu77},
\beqs
z_{sq} = \frac{4C}{\pi} & = & \frac{1}{2} \int_{-\pi}^{\pi} \frac{d\theta_1}{2\pi} \int_{-\pi}^{\pi} \frac{d\theta_2}{2\pi}
\ln \Big [ D_{sq,2}(\theta_1,\theta_2) \Big ] \cr\cr
& = & \frac{1}{3} \int_{-\pi}^{\pi} \frac{d\theta_1}{2\pi} \int_{-\pi}^{\pi} \frac{d\theta_2}{2\pi}
\ln \Big [ D_{sq,3}(\theta_1,\theta_2) \Big ] \cr\cr
& = & \frac{1}{4} \int_{-\pi}^{\pi} \frac{d\theta_1}{2\pi} \int_{-\pi}^{\pi} \frac{d\theta_2}{2\pi}
\ln \Big [ D_{sq,4}(\theta_1,\theta_2) \Big ] \ .
\eeqs
The same procedure can be carried on for lattices with different choices of unit cells, and many integral identities can be obtained.

\section{Homeomorphic Expansion}

\begin{figure}[htbp]
\unitlength 1mm \hspace*{5mm}
\begin{picture}(60,60)
\multiput(0,20)(0,20){2}{\line(1,0){60}}
\multiput(20,0)(20,0){2}{\line(0,1){60}}
\multiput(10,20)(10,0){5}{\circle*{2}}
\multiput(10,40)(10,0){5}{\circle*{2}}
\multiput(20,10)(0,10){5}{\circle*{2}}
\multiput(40,10)(0,10){5}{\circle*{2}}
\put(22,32){\makebox(0,0){\small 1}}
\put(22,22){\makebox(0,0){\small 2}}
\put(32,22){\makebox(0,0){\small 3}}
{\linethickness{1mm}
\multiput(20,0)(0,20){3}{\line(0,1){10}} 
\multiput(40,0)(0,20){3}{\line(0,1){10}}
\multiput(0,20)(20,0){3}{\line(1,0){10}} 
\multiput(0,40)(20,0){3}{\line(1,0){10}} }
\end{picture}

\vspace*{5mm}
\caption{\footnotesize{$sq(1)$, the homeomorphic expansion of the square lattice with $p=1$ vertex inserted on each edge. Vertices within a unit cell are labeled and edges within each unit cell are connected by thick lines.}} 
\label{sqhe}
\end{figure}
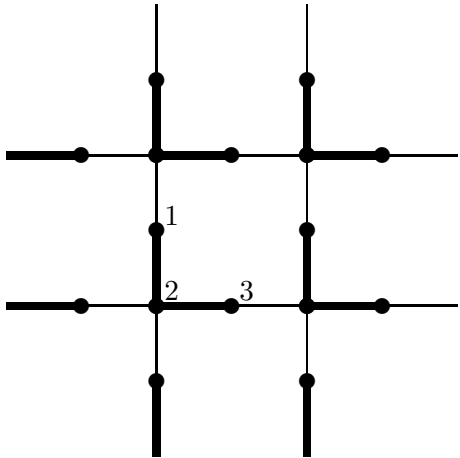

The method reviewed in section \ref{sectionmethod} can be applied to lattices which are not $k$-regular. We find that for a $k$-regular lattice $\Lambda$, the asymptotic growth constant of its homeomorphic expansion $\Lambda(p)$ is given by
\beq
z_{\Lambda (p)} = \frac{(\frac\kappa 2-1) \ln (p+1)+z_\Lambda}{\frac\kappa 2 p+1} \ .
\label{zlambdap}
\eeq
This can be understood as follows. As the number of occupied edges by a spanning tree on the original lattice is $n-1$, the number of unoccupied edges on the original lattice is $e(G)-(n-1) = (k/2)n-n+1$. Now on each unoccupied edge we insert $p$ degree-2 vertices. There are $p+1$ ways to make these vertices connected in order to construct spanning trees on the homeomorphic expanded lattices, so that
\beq
N_{ST}(\Lambda(p)) = (p+1)^{\frac k2 n-n+1} N_{ST}(\Lambda) \ .
\eeq
Eq. (\ref{zlambdap}) follows by noticing that the total number of vertices of the homeomorphic expanded lattice is $(k/2)np+n$. For the homeomorphic expansion of the square, triangular and honeycomb lattices, 
\beqs
z_{sq(p)} & = & \frac{\ln (p+1)+z_{sq}}{2p+1} \cr\cr
z_{tri(p)} & = & \frac{2\ln (p+1)+z_{tri}}{3p+1} \cr\cr
z_{hc(p)} & = & \frac{\frac12 \ln (p+1)+z_{hc}}{\frac32p+1}
\eeqs
where $p$ is a positive integer. For a simple illustration, consider the square lattice with one vertex inserted on each edge. As shown in Fig. \ref{sqhe}, the number of vertices in each unit cell $\nu_{sq(1)}$ is equal to three. We have
\beq
M_{sq(1)} =
\left(\begin{array}{ccc}
2 & -1-e^{i\theta_2} & 0 \\
-1-e^{-i\theta_2} & 4 & -1-e^{-i\theta_1}  \\
0 & -1-e^{i\theta_1} & 2 
\end{array} \right) \ .
\label{Msqh2}
\eeq
The determinant $D_{sq(1)}$ is equal to $8-4\cos \theta_1-4\cos \theta_2$, which is twice of the corresponding expression for the regular square lattice \cite{sw}, such that $z_{sq(1)}=(\ln 2 +z_{sq})/3$. 

\section{Discussion}

We summarize the values of $z_\Lambda$ and $r_\Lambda$ for various lattices $\Lambda$ in Table \ref{ztable}. Our results agree with the observations made in Ref. \cite{sw}. Namely, $z_\Lambda$ is relatively large for large value of $k$; while for a fixed $k$ value, $z_\Lambda$ increases with the spatial dimension of the lattice. An example is given by the circuit graph with next-nearest-neighbor edges, $(3.4.6.4)$ lattice and the diamond structure (all with $k=4$). The ratio $r_\Lambda$ increases with dimension and approaches to one.

\begin{table}
\caption{\label{ztable} Values of $z_\Lambda$ and $r_\Lambda$.}
\footnotesize\rm
\begin{tabular*}{\textwidth}{@{}l*{15}{@{\extracolsep{0pt plus12pt}}l}}
\br
$\Lambda$ & $d$ & $k$ & $z_\Lambda$ & $r_\Lambda$ \\
\mr
$(3^3.4^2)$   & 2 &  5 &  1.406925832  &  0.9486371781 \\ 
$(3^2.4.3.4)$ & 2 &  5 &  1.410855646  &  0.9512869040 \\ 
$(3^4.6)$     & 2 &  5 &  1.392023563  &  0.9385891391 \\ 
$(3.4.6.4)$   & 2 &  4 &  1.144801124  &  0.9411423250 \\ 
$(4.6.12)$    & 2 &  3 &  0.7777955061 &  0.9292789200 \\ 
$c(1),nnn$    & 1 &  4 &  0.9624236501 &  0.7912095935 \\ 
$c(2),nnn$    & 2 &  8 &  1.943739360  &  0.9681095375 \\ 
$c(3),nnn$    & 3 & 18 &  2.841787536  &  0.9933044903 \\ 
$c(4),nnn$    & 4 & 32 &  3.442108     &  0.9978270    \\ 
$c(5),nnn$    & 5 & 50 &  3.898251     &  0.9990858    \\ 
diamond       & 3 &  4 &  1.206459950  &  0.9918321170 \\ 
\br
\end{tabular*}
\end{table}

\section{Acknowledgments}
We thanks C.-H. Chen for use of computer facilities and software.
S.-C. C. thanks Prof. R. Shrock for helpful discussion and the collaboration on Ref. \cite{std}. This research was partially supported by the Taiwan NSC grant NSC-94-2112-M-006-013.

\end{document}